\documentclass{icontop}

\begin{document}

\title{Development of ARIES Baker-Nunn camera to a wide-field Imaging Telescope with CCD}

\author{Soumen Mondal, K. G. Gupta, Sneh Lata, Biman J. Medhi, Tarun Bangia, T. S. Kumar, Shobhit Yadav \& S.K. Singh}

\institute{Aryabhatta Research Institute of Observational Sciences (ARIES),\\ 
Manora Peak, Nainital -263 129, India \\ email: soumen@aries.ernet.in }
 
\maketitle

\begin{abstract}

ARIES Baker-Nunn Schmidt telescope project is converting a Baker-Nunn satellite tracking camera for Astronomical research. Original Baker-Nunn camera produces an extremely large (5$\times$30 $degree$) curved focal plane at the prime focus for photographic imaging. We present here the re-designing of the camera  produces  a wide (4$\times$4 $degree$) flat field of view for CCD imaging observations, which have many scientific potentials in Astronomy. Imaging performance of the CCD camera is also estimated.
 
\end{abstract}

\section{Introduction}

The ARIES Baker-Nunn Schmidt telescope project was initiated during 2005 to use
its' wide-area imaging capabilities for numerous scientific purposes with modern CCD detector. The telescope will be installed at Manora Peak, Nainital, India (longitude: 79$^o$ 18$^{\prime}$ 00$^{\prime\prime}$ E  and latitude: 29$^o$ 22$^{\prime}$ 00$^{\prime\prime}$ N). 

Scientific programs like study of variable stars, Asteroids and
Near-Earth-Objects, detecting of extra-solar planets though transit method,
transient objects like GRBs and supernovae, and imaging of large star clusters
could be suitably accomplished with this wide-field imaging telescope. The
photographic Palomar all-sky surveys  is one of most important contribution from
the 1.2-m Samuel Oschin Schmidt, 1.2m UK-Schmidt and 1-m ESO Schmidt telescopes
\cite{reid}.  The Near-Earth Asteroid Tracking (NEAT)\footnote{http://neat.jpl.nasa.gov/}
project has recently instrumented the Palomar Samuel Oschin telescope with 4080
$\times$ 4080 CCDs covering of $\sim$3.75 deg$^2$ field of view \cite{pravdo}. Recently, QUEST (Quasar Equatorial Survey Team) collaboration \footnote{The collaboration includes several US Universities and CIDA (Venezuela)} uses the 1-m Schmidt telescope located at the Venezuelan National Astronomical Observatory with 16 CCDs (2048 $\times $2048) arranged in a 4
$\times$ 4 mosaic covering 2.3$^o$ $\times$ 3.5$^o$ of the sky \cite{baltay}. The Automated Patrol Telescope (APT)\footnote{http://www.phys.unsw.edu.au/~mcba/apt/} at
Siding Spring Observatory, Australia operated by the University of New South
Wales (UNSW) uses a 0.5m telescope with 770 $\times$ 1150 CCD covering 2$^o$
$\times$ 3$^o$ field of view \cite{carter1,carter2}. The UNSW
Extra-solar Planet Search is the major scientific project using the APT facilities \cite{hidas}, and produced several variable stars as a byproducts \cite{chris1,chris2}. The Fabra-ROA telescope\footnote{http://www.am.ub.es/bnc/} project aims to refurbish an f/1 0.5m Baker-Nunn Camera (BNC) with
4096 $\times$ 4096 CCD covering a field of view of 4.4$^o$ $\times$ 4.4$^o$  for
the robotic surveying purposes, this re-furbishing project is  similar like the
ARIES BN Schmidt telescope project. 

The basic optical design of the Baker-Nunn camera at ARIES  uses a
three-element corrector to produce an extremely wide field of view ( 5$^o$
$\times$ 30$^o$) across a curved plane at the prime focus for photographic
imaging.  The transformation of the Baker-Nunn camera  into a 50 cm Schmidt
telescope with CCD imaging capabilities is a process which comprises several
steps in different categories.  Major jobs in this project are  (i) modification
of optical design for curved to flat focal plane (CCD observations), (ii)
changing the mounting system from alt-azimuthal to Equatorial english mount  (mechanical
re-designing), (iii) computer control of the telescope, (iv) optical alignment
and installing new customize CCD imaging system.  

\section{Optics re-designing}

To accommodate CCD detectors in the focal plane, 
the field should be flat. The optical design of the telescope needs to be
corrected for flat focal plane as wide  as possible with acceptable geometric
distortion. The  new design for Baker-Nunn
Schmidt telescope incorporates two new optical elements, one field flattener
lens and one meniscus lens, with old optical elements that include 80 cm f/1
primary mirror and 50 cm aperture 3-elements corrector lens system. 
The field flattener is used to flatten the curved field. The lens is  55 mm
diameter double convex lens and placed inside CCD camera at distance 1.66 mm 
from the CCD chip to avoid from introducing unacceptable
aberrations. A meniscus lens is placed between focal plane and primary mirror. The meniscus lens provides the correction for the astigmatism
introduced by the field flattener.
The optical
layout of the BNC telescope   showing new optical elements is shown in Fig.~1a. 

\begin{figure}[htb]
\includegraphics[width=5.5cm]{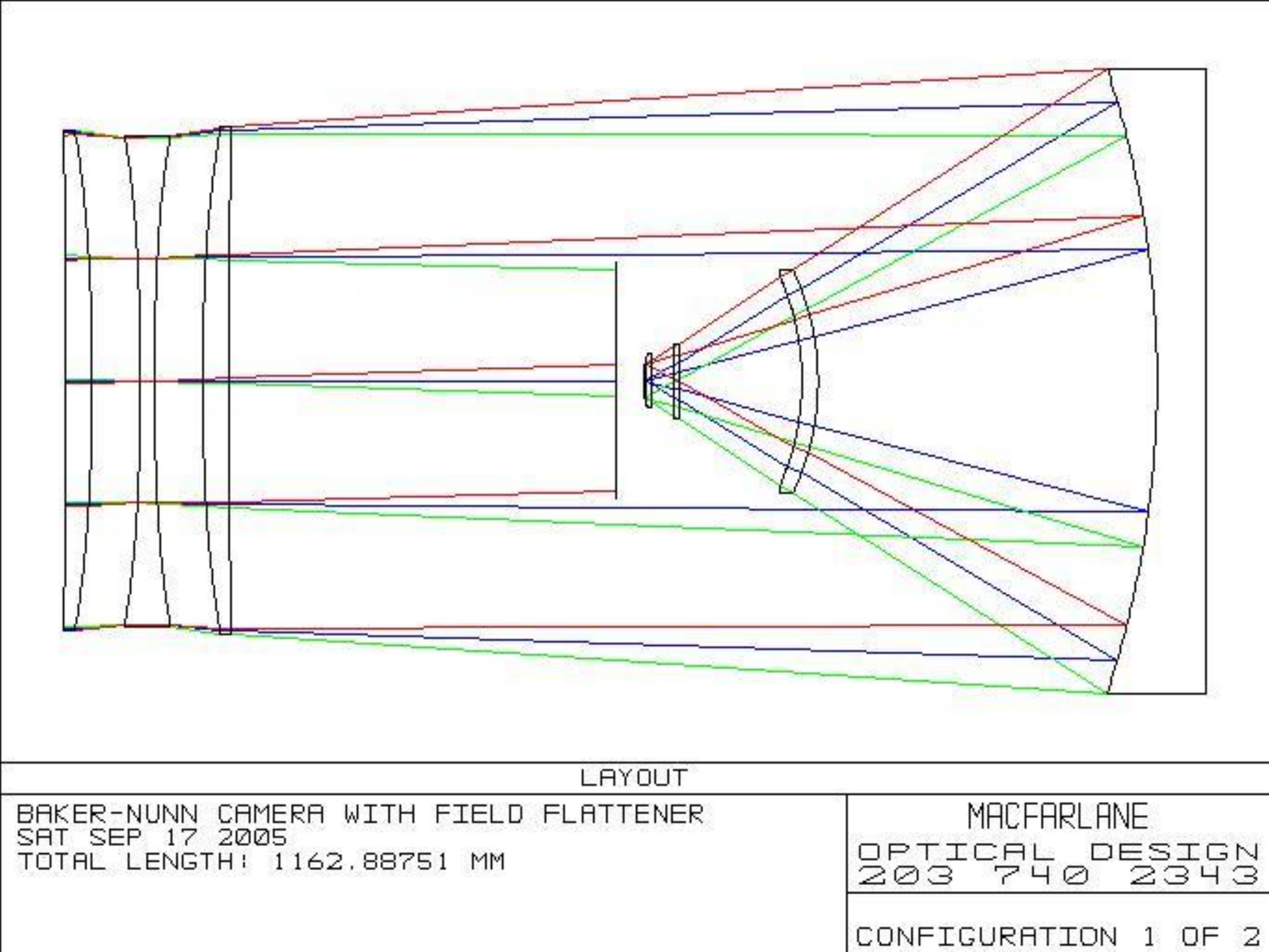}\hspace{0.1in}
\includegraphics[width=5.5cm]{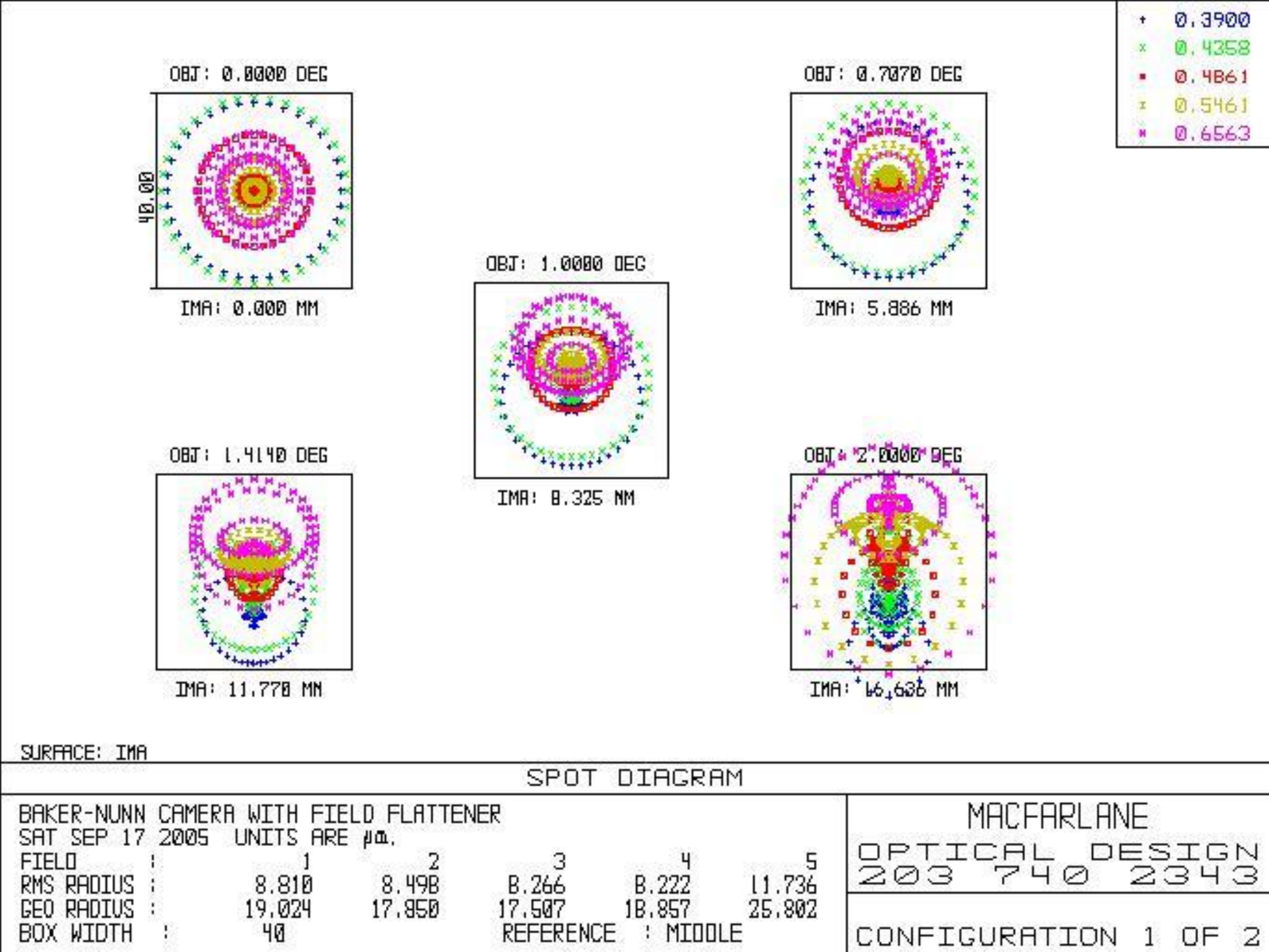}
\caption[Optical layout of BNC]{Fig.~1a: the optical layout of the ARIES Baker-Nunn
telescope. Fig.~1b: shows spot diagrams at
5 semi-field locations 0.0, 0.707, 1.00, 1.414, 2.00 degrees.}
\end{figure}

The optical system of the Schmidt telescope consists of a 3-element air-spaced
Schmidt corrector lens system  with an aperture of  50-cm, a spherical primary
mirror of 80 cm  diameter. The optical elements and their positions in the telescope including some useful parameters are listed in the table 1.

\begin{table}[htb]
\begin{center}
\caption{Optical elements, their positions, telescope parameters}
\vskip 0.15in
\begin{tabular}{l|l}
\hline
\hline
Optical element & Position w.r.t. the Primary mirror/parameters\\
\hline
Primary mirror &  0 $mm$ \\
Meniscus lens  & 345 $mm$  \\
Field-Flattener lens & 520.11 $mm$ \\
3-element Corrector lens system &  1110.57 $mm$ \\
Focal surface & 521.77 $mm$ \\
Telescope F-number & 0.9539 \\
Telescope plate scale & 412 $arcsecond/mm$ \\
Radius of imaging area & 18 $mm$ \\
\hline
\end{tabular}
\end{center}
\end{table}

\section{Mechanical re-design}

Schmidt Telescope has a spherical primary mirror of diameter 80 cm housed in primary mirror cell supported on 18 axial and 12 radial supports, which are balanced to support exactly the proportion of weight of the mirror coming on them at any position of the primary mirror. Primary Mirror cell is mounted at the bottom end of the tube. Baker Nunn corrector and the prisms are mounted on the front (top end) of the tube. They can be adjusted to align with the axis of the primary. Tube houses the CCD Assembly, meniscus assembly and declination counter weight assembly. Two guide telescopes are mounted outside the tube in diametrically opposite locations.  These are mounted on adjustable mounts for aligning their axes with the axis of the tube (primary). The tube assembly is fixed to the declination trunnion, which rotates in the central section of the mount. Mount assembly supports the tube.  Mount contains the declination and polar axes about which the telescope rotates to locate and track the objects in the sky. The  mount consists of the declination axis assembly, south bearing assembly, north bearing assembly, declination drive assembly, R.A. counterweight assembly and R.A. drive assembly. The telescope about polar axis and the tube assembly about the declination axis are rotated by identical gearboxes.  The gearboxes are shaft mounted and fitted with torque arms to provide torsional rigidity. Gearbox is mounted on the bearings fitted on the bull gear.

\section{Performance of the optical design}

The optical performance were studied by optimizing all optical elements
in the telescope and the operating wavelengths using the $Zemax$. Fig.~1b shows  the spot
diagrams at the five semi-field points located at 0.00,  0.707, 1.00, 1.414 and
2.00 degrees from the center in the visible wavelength. The spot sizes are in the range of 35 to 50 $\mu$m over the entire field. The encircled energy (EE) of 90\%  is expected to lie within 30 $\mu$m diameter region up to 1.414 degrees of the semi-field, while 80\% at the edges. 
Optical analysis shows that the field curvature would be less than 20 $\mu$m at the edges, and the distortion of about 0.09\% at the edges.  For a 18 mm semi-field size at the focal plane, a  distortion of 0.09\% corresponds to 16 $\mu$m.

\section {CCD camera system}

The CCD camera is located at the prime focus of ARIES Baker-Nunn Schmidt
telescope, inside the telescope tube at about 521.77 mm from the primary mirror.
The f/1 fast optics covers 4  degree  circular field of view with the meniscus
lens placed between CCD detector and primary mirror at 345 mm from the primary.
To make a change from the curved to flat image plane, the modified 
optical design incorporates a field-flattener lens  close to (in our system, it is 1.66 mm) the CCD chip. As the CCD camera sits inside the telescope tube at the {\it prime focus}, the
outside dimensions of the camera have to be compact for minimum obscuration of
the incoming beam from the primary.

The CCD detector is Kodak KAF-168003 chip with 4096$\times$4096 pixels, each 9
$\mu$m size.  The detector is used in the front-illuminated mode, the quantum
efficiency (QE) is about 50-59\% in the visible band. Technical specifications are shown in the table 2.
 
\begin{table}[thb]
\begin{center}
\caption{Technical specifications of the CCD camera}
\vskip 0.1in   
\begin{tabular}{l|l}
\hline
\hline
CCD chip & Kodak KAF-168003 \\
Quantum efficiency &  $\sim$55 \% at V \\
Array size & 4096 $\times$ 4096 \\
Pixel size & 9 $\mu$m \\
Imaging area & 36.68 $\times$ 36.86 $mm$ \\
Read Noise & 9-11 $electrons$ \\
Dark current & $<$ 1 $electron/pixel/sec$ (at -30$^o$ C) \\
Operating Environment & temperature: -25$^o$ to 35$^o$ C. Relative humidity 10 to
90\% \\
Cooling system & Thermoelectric cooler with fan-assisted air cooling; \\
               & operates 55-60 deg. C below the ambient temperature; and \\
               & also the optional exchangeable liquid cooling  \\ \\ 
Linear Full well & 100K $electrons$ \\

              &  \\
Camera outer dimension & 6.2 $inch$ width $\times$ 6.2 $inch$ depth $\times$  4.2
$inch$ height \\ 
Camera weight & $\sim$ 6 $pound$ \\
\hline 
\end{tabular}
\end{center}
\end{table}

\section{Opto-mechnical alignment}

The optical alignment of a Schmidt telescope requires the fulfillment of three
simultaneous conditions: (1) to obtain the coincidence of the optical axis of
the corrector plate with the geometrical axis of the tube, (2) to locate the
center of curvature of the spherical mirror at center of the corrector plate,
(3) to orient the detector plane (centering and tilting) according to its focal
surface. 

The following proposed alignment procedures could be followed  
as described by Anderson \& Clausen (1974) \cite{andersen} :

\noindent
i. Proper placing of the primary mirror in the cell. \\
ii. Centering of the corrector elements. \\
iii. Removing of the detector tilt. \\
iv. Centering of the detector. \\

\begin{table}
\begin{center}
\caption{ Estimated throughput for the ARIES B.N. Schmidt telescope}
\vskip 0.1in
\begin{tabular}{l|l|l}
\hline
\hline
Optical components & Transmission & Cumulative efficiency \\
\hline
3-element corrector & 0.9$\times$0.9$\times$0.9 & 0.729 \\
Primary mirror    & 0.95  &   0.693 \\
Meniscus lens & 0.9 & 0.623 \\
Field Flattener lens & 0.9 & 0.561 \\
Filters (V,R)  & 0.8 & 0.448 \\
CCD QE & 0.55 & 0.246 \\
\hline
{\bf Total} & & {\bf 0.246} \\
\hline
\end{tabular}
\end{center}
\end{table}

\begin{table}[htb]
\begin{center}
\caption[Parameters for telescope and detector]{Parameters used for SNR calculation}
\vskip 0.1in
\begin{tabular}{l|l}
\hline
\hline
Telescope diameter & 80 $cm$ \\
Clear aperture & 50 $cm$  \\
Plate scale   & 412 $arcsecond/mm$  \\
Readout noise of the detector & 10 $electrons/ADU$  \\
Dark current of the detector  & 1 $electron/pixel/sec$ \\
Pixel size & 9 $\mu$m \\ 
Saturation limit & 100000 $electrons$ \\
Thoughput of the system & 0.246   \\
Sky brightness in V & 21.8 $mag/arcsecond^2$ \\  
\hline
\end{tabular}
\end{center}
\end{table}

\section {Detection limit with the CCD}

For a CCD, the signal to noise ratio (SNR) from a point source  is determined by
a well-known CCD equation \cite{howell}, which is given by,

\begin{equation}
S/N=\frac{N_{star}}{\sqrt{N_{star} + n_{pix}(N_{sky}+ N_{dark}+ N_{read}^2)}} 
\end{equation}

Where N$_{star}$ is the total photon counts from the star (sky subtracted),
N$_{sky}$ is counts from the sky. The N$_{dark}$ and N$_{read}$ are the dark
current and read out noise, and it is the characteristic of the CCD. The
n$_{pix}$ is the total number of pixels contained within the selected aperture
for the photometry, it is related to {\it seeing} of the site. Here, star images
are approximately 20 -25 $\mu$m in diameter, this spot size is a practical lower
limit for the optical system.  Considering the image size, n$_{pix}$ is chosen
to be $\sim$ 3 pixels, corresponding  to $\sim$ 11 {\it arcsecond} in the sky.
The {\it seeing} is $\sim$ 2 -2.5 {\it arcsecond} for our site, which is far
below the spot size of an image.   

The parameters, listed in the table 3 and 4, are considered for
estimation of the SNR and shown in Fig.~2. The SNR is basically related to the photometric error,
$\sigma_m$ $\sim$ {\it 1.085/SNR}. The predicted magnitude limit is V $\approx$
19 for a 10$\sigma$ detection and an exposure time of 300 $sec$. A photometric
precision of 1\% ($\sigma_{mag}$ $\sim$ 0.01) for a V = 15.5 star with exposure
time of 100 $sec$ is estimated from photon statistics from the star and the sky
background.  The theoretical limit of the maximum photometric accuracy from the
system is related to the linear full well capacity of the detector, the limit
here is 3 {\it millimagnitude}.

\begin{figure}[htb]
\includegraphics[width=10cm]{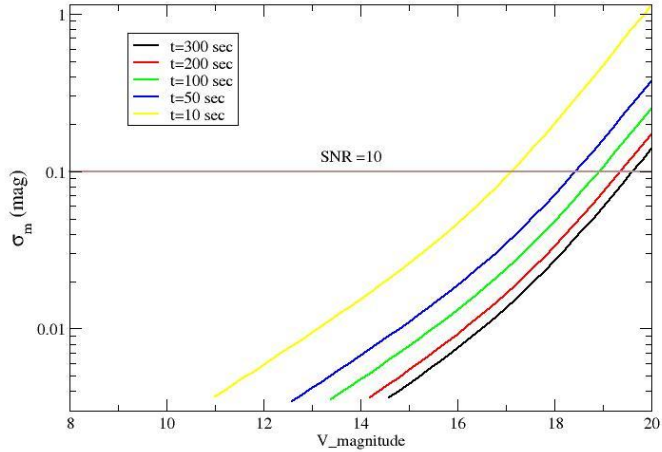}
\caption[V magnitude versus photometric precision]{The graph shows the
predicted photometric precisions versus V magnitude for the
different exposure times considering the parameters in the table 4. The
horizontal line shows the signal to noise ratio (SNR=10) for the different exposure times. The bright
end of those curves limit up to the saturation limit of the detector.}
\end{figure} 

\section{Summary}

We present here the development of ARIES Baker-Nunn telescope project, which is re-designed from a Baker-Nunn satellite tracking camera. The telescope will be used in a wide-field imaging mode using the custom designed CCD camera for various Astronomical research programs. The optical components in the new design, like field-flatterner lens and minicus lens, are procured from the standard optical companies.  The custom designed CCD camera is in the processes of the procurement from the Finger Lakes Instrument, USA. The first-light from the telescope is expecting during the mid-2009.

\vspace{10pt}
\noindent Acknowledgements: The authors are thankful to Dr. Malcolm J.
MacFarlane for the optical re-design. The authors are extremely thankful to Prof. Ram Sagar, Director, ARIES for his constant support and encouragement for this project.


\begin{thebibliography}{99}

\bibitem{reid} N. Reid, et al. ``The Second Palomar Sky Survey'' Pub. Astron. Soc. Pacific {\bf 101} (1991)  661-674  

\bibitem{pravdo} S.H. Pravdo, et al. ``The Near-Earth Asteroid Tracking (NEAT) Program: an Automated System for Telescope Control, Wide-Field Imaging, and Object Detection'' The Astron. Journal {\bf 117} (1999) 1616-1633 

\bibitem{baltay} C. Baltay et al. ``A Large-Area CCD Camera for the Schmidt Telescope at the Venezuelan National Astronomical Observatory'' Pub. Astron. Soc. Pacific {\bf 114} (2002) 780-794 

\bibitem{carter1} B.D. Carter, M.C.B. Ashley, Y.-S. Sun \& J.W.V Storey ``Redesigning  a Baker-Nuun Camera for CCD Imaging'' 1992, {\it Proc. Astron. Soc. Austrilia}, 10, 74-76  

\bibitem{carter2} B.D. Carter, et al. ``Astronomy with the Automated Patrol Telescope'' ASP conf. Series {\bf 5} (1995)  44-47  

\bibitem{hidas} M.G. Hidas, et al. ``The University of New South Wales Extrasolar Planet Search: methods and first results from a field centred on NGC 6633'' Mon. Not. Royal Astron. Soc. {\bf 360} (2005) 703-717 

\bibitem{chris1} J.L. Christiansen, et al. ``The first high-amplitude δ Scuti star in an eclipsing binary system''  Mon. Not. Royal Astron. Soc. {\bf 382} (2007) 239-244  

\bibitem{chris2} J.L. Christiansen, et al. ``The University of New South Wales Extrasolar Planet Search: a catalogue of variable stars from fields observed between 2004 and 2007'' Mon. Not. Royal Astron. Soc. {\bf 385} (2008) 1749-1763

\bibitem{andersen} J. Andersen, J. \& J.V. Clausen `` Adjustment and testing of Schmidt telescopes'' Astron. \& Astrophys. {\bf 34}  (1974) 423-429  

\bibitem{howell} S.B. Howell ``Two-dimensional aperture photometry - Signal-to-noise ratio of point-source observations and optimal data-extraction techniques'' Pub. Astron. Soc. Pacific {\bf 101} (1989)  615-622  

\end{thebibliography}
\end{document}